*Type of the Paper (Article)*

# An Investigation of Hepatitis B Virus Genome using Markov Models

**Khadijeh (Hoda) Jahanian[#1], Elnaz Shalbafian[#2], Morteza Saberi[1], Roohallah Alizadehsani[3], Iman Dehzangi[4,5]**

[1]Faculty of Engineering and IT, University of Technology Sydney, Sydney, AU

[2] Isfahan University of Technology, Isfahan, IR

[3]Institute for Intelligent Systems Research and Innovation (IISRI), Deakin University, Victoria, AU

[4]Department of Computer Science, Rutgers University, Camden, NJ, 08102, USA

[5]Center for Computational and Integrative Biology, Rutgers University, Camden, NJ, 08102, USA

# Joint first author

* Corresponding authors: K.J.

**Abstract:** The human genome encodes a family of editing enzymes known as APOBEC3 (apolipoprotein B mRNA editing enzyme, catalytic polypeptide-like 3). Several family members, such as APOBEC3G, APOBEC3F, and APOBEC3H haplotype II, exhibit activity against viruses such as HIV. These enzymes induce C-to-U mutations in the negative strand of viral genomes, resulting in multiple G-to-A changes, commonly referred to as 'hypermutation.' Mutations catalyzed by these enzymes are sequence context-dependent in the HIV genome; for instance, APOBEC3G preferentially mutates G within GG, TGG, and TGGG contexts, while other members mutate G within GA, TGA, and TGAA contexts. However, the same sequence context has not been explored in relation to these enzymes and HBV. In this study, our objective is to identify the mutational footprint of APOBEC3 enzymes in the HBV genome. To achieve this, we employ a multivariable data analytics technique to investigate motif preferences and potential sequence hierarchies of mutation by APOBEC3 enzymes using full genome HBV sequences from a diverse range of naturally infected patients. This approach allows us to distinguish between normal and hypermutated sequences based on the representation of mono- to tetra-nucleotide motifs. Additionally, we aim to identify motifs associated with hypermutation induced by different APOBEC3 enzymes in HBV genomes. Our analyses reveal that either APOBEC3 enzymes are not active against HBV, or the induction of G-to-A mutations by these enzymes is not sequence context-dependent in the HBV genome.

**Keywords:** APOBEC3 enzymes; Clustering; PCA; Markov models; Multivariate data analytics; Motif analysis.

## 1. Introduction

The APOBEC3 gene, located on human chromosome 22 [1], comprises at least seven genes. APOBEC3 family members function as C-to-U editing enzymes, playing crucial roles in the antiviral response. These enzymes catalyze cytidine deamination in single-stranded DNA (ssDNA) through a catalytic center formed by a histidine and two cysteines [2]. During deamination, cytidine is replaced by a carbonyl, resulting in the transformation of cytidine to uridine (C-to-U). Extensive research has shown that APOBEC3 enzymes inhibit certain viruses by inducing C-to-U mutations in the minus strand of the viral DNA, leading to multiple G-to-A replacements, commonly termed 'hypermutation,' in the viral positive strand [3]. Among these enzymes, APOBEC3G and APOBEC3nonG (other members of the APOBEC3 family) have been reported to induce G-to-A changes in the HIV-positive strand within specific target motifs. Specifically, APOBEC3G preferentially replaces G with A within a GG context, while other family members mainly induce G-to-A changes within GA contexts [4].

To date, the relationship between these enzymes and HBV has not been thoroughly explored. This study aims to identify the motif preferences of APOBEC3G and APOBEC3nonG enzymes in the HBV genome.

Since the discovery of APOBEC3G (A3G) as a restriction factor, significant progress has been made in understanding its mechanism of action against the pararetrovirus HBV. Turelli et al. [5] reported that the inhibition of viral pregenome packaging RNA is the most probable target of A3G, and A3G can inhibit DNA replication in vitro related to HBV. Subsequently, Rösler et al. [6] confirmed the inhibition of pregenome packaging RNA in duck HBV in vitro. Although clear evidence for viral DNA degradation was not identified, they suggested that APOBEC3G inhibited DNA-RNA hybrid replication in HBV [7].

In 2007, Bonvin et al. [8] observed the expression of APOBEC3 cytidine deaminases in the normal human liver and reported low mRNA expression of these proteins. Their investigations demonstrated that dual deaminase domain enzymes, A3G, A3F, and A3B, can edit DNA replication in HBV, generating extensively G-to-A hypermutated HBV genomes [8]. Later, Noguchi et al. [9] confirmed the complete inhibition of Hepatitis B virus replication in vitro, reporting that the expression of A3G and A3F increased the number of hypermutated genomes while decreasing the number of HBV replications [9]. Concurrently, Bonvin and Greeve [10] employed A3B as a model enzyme in vitro to study the inhibition of HBV replication associated with catalytic cytidine deaminase activity. They reported that the induction of G-to-A hypermutations in HBV alone is insufficient to fully inhibit HBV replication.

However, detecting APOBEC3-induced mutations in hypermutated viral sequences typically involves aligning sequences and comparing them to a reference sequence. Numerous studies have analyzed hypermutated sequences to identify APOBEC3G and APOBEC3nonG target motifs using alignment to a constructed reference [4, 11]. While alignment is useful when a parental reference sequence is available, it may not be suitable in cases where the parental sequence is unavailable, as is often the case with individuals infected with natural HBV, HIV, or SIV. Additionally, indels may occur in alignment-based methods, leading to the inaccurate classification of APOBEC3 target and non-target sites. This results in an incorrect assignment of the frequency of GG-to-AG and GA-to-AA mutations [12, 13]. Therefore, it is crucial to develop a method capable of identifying and quantifying hypermutated sequences without the need for alignment to a reference sequence.

The application of machine learning methods [14-17] and statistical approaches [18-20] in medical data analytics including genomic motif analysis facilitates the identification and interpretation of intricate patterns within DNA sequences, enhancing our understanding of genetic regulation and contributing to the discovery of novel functional elements in the genome. Researchers have recently demonstrated that employing Markov Models (MM) of conditional probabilities provides a more accurate estimation of motif representation [16, 21, 22]. In these approaches, the expected frequency of a motif is estimated by considering the observed frequencies of the motif constituents, accounting for overlapping nucleotide(s) [21, 22].

Motifs such as GG and GGG, which are preferentially targeted and mutated by APOBEC3, exhibit lower representation in the genome of hypermutated sequences compared to non-hypermutated sequences. Conversely, for product motifs such as AG and AGG, the opposite trend is observed. Consequently, the examination of motif representation serves as a valuable tool for investigating hypermutation. In this study, we quantify the representation of short sequence motifs and employ multivariate analysis to discern the motif preferences of APOBEC3 proteins.

In this study, we conducted a comprehensive investigation into the representation of all K-mer motifs (genomic motifs with a length of k, e.g., ACG in a 3-mers motif or ACGT in a 4-mers motif), totaling 340, where k ranges from 1 to 4 (A, C… AA, AC… TTTT). This analysis aimed to identify underrepresented (targeted motifs) or overrepresented (produced motifs) motifs in the genome of hypermutated HBV sequences compared to normal (non-hypermutated) HBV sequences. Utilizing 901 Human HBV sequences from Vartanian et al. [23], which include numerous hypermutated sequences, we computed the representation (also referred to as D-ratio) of all 340 K-mer motifs in these genomes. To identify motifs associated with hypermutation, we employed an analytical approach that elucidates the relationship between HBV sequences and motifs.

The representation data of the 340 K-mer motifs in these HBV sequences form a matrix, where K-mers and HBV sequences serve as variables and objects, respectively (Fig. 1). Each HBV sequence is defined by 340 variables, and each K-mer is characterized by the HBV sequences. Exploratory analysis of such a data matrix necessitates a multivariate approach, such as Principal Component Analysis (PCA). This method reveals differences among HBV subtypes regarding the representation of K-mer motifs and discrepancies among K-mer motifs within the HBV sequences. Most importantly, it facilitates the identification of K-mer motifs that describe similarities and dissimilarities among HBV sequences and vice versa.

## 2. Experimental Section

*HBV sequences*

We used 3,000 human HBV normal sequences downloaded from NCBI and 901 HBV hypermutated sequences from Vartanian *et al*. [23].

*HIV-1 sequences*

We used 2,047 full genome HIV-1 sequences containing 54 hypermutated viral sequences from the subtypes A1, B, C, and the recombinant 01_AE, from naturally infected patients that were obtained from the Los Alamos National Laboratory (LANL) database. The accession numbers of hypermutated sequences used in this study are listed in **Table 1**.

**Table 1.** Accession number of hypermutated sequences from the Los Alamos National Laboratory (LANL) database.

| | Subtype | Accession numbers |
|---|---|---|
| 1 | A1 | EF165366; EF165365; AF484484; AF457091; AF457076; AF457071; AF457057; FJ388907. |
| 2 | B | EF165363; AY829213; AY037274; AY779556; AY781125; AY818643; AY818642; AY818641; AY531116; AY561241; FJ195087; FJ388922; FJ388897; JF689891; JF689888; JF689882; JF689881; JF689880; JF689878; JF689861; JF689858; JF689855; JN235961; EF178404 |
| 3 | C | EF165360; EF165359; DQ275665; DQ164128; DQ164125; DQ164124; DQ164123; DQ056407; AY734561; AY734557; AY255828. |
| 4 | 01_AE | EF165361; AY945729; AY945723; AY945715; AY945714; AY358058; AY358055; AY358054; AY358053; GU201515; GU564226 |

*Markov model*

We define "representation" as the ratio of observed frequency ($P_{obs}$) of a motif over its expected frequency ($P_{exp}$) in the genome. The $P_{obs}$ of a motif is defined as the number of times that motif appears in the sequence divided by the total number of all motifs with the same length. The $P_{exp}$ can be calculated in different ways.

$$P_{exp}(CpGpT) = P_{obs}(C) \times P_{obs}(G) \times P_{obs}(T) \quad \text{Eq. 1}$$

$$= P_{obs}(CpG) \times P_{obs}(T) \quad \text{Eq. 2}$$

$$= P_{obs}(C) \times P_{obs}(GpT) \quad \text{Eq. 3}$$

In this study, we use 1st and 2nd order models to calculate the expected frequencies of tri-nucleotides and tetra-nucleotides, respectively. Examples of 1st and 2nd order Markov models are given in Equations 4 and 5.

$$P_{exp}(CpGpT) = \frac{P_{obs}(CpG) \times P_{obs}(GpT)}{P_{obs}(G)} \quad \text{Eq. 4}$$

$$P_{exp}(CpGpTpA) = \frac{P_{obs}(CpGpT) \times P_{obs}(GpTpA)}{P_{obs}(GpT)} \quad \text{Eq. 5}$$

We then calculate its representation (D-ratio) for each motif by dividing the $P_{obs}$ of the motif by its $P_{exp}$. This is shown using an example in Eq. 6. The expected probability of the tetra-nucleotide motif TGGG is computed using the observed probabilities of its dinucleotides (TG, GG) and mononucleotide G constituents.

$$D(TGGG) = \frac{p_{obs}(TGGG)}{p_{exp}(TGGG)} = \frac{p_{obs}(TGGG)}{\frac{p_{obs}(TGG) \times p_{obs}(GGG)}{p_{obs}(GG)}} \quad \text{Eq. 6}$$

The computed D-ratio in Eq. 6 serves as a pure 'representation' of TGGG, and notably, it remains independent of potential changes in TGG, GGG, and GG. Employing Markov Models (MM), we generate an extensive array of features to articulate our data. Handling a large number of variables in data, as encountered in high-dimensional datasets such as biological datasets, can pose challenges in extracting meaningful information. Therefore, a reduction in the number of variables is essential without compromising useful information. In this context, we utilize the PCA model to address this issue, elucidated in detail in the following subsection.

Principal Component Analysis

PCA stands as a robust statistical technique employed to transform a set of observations involving possibly correlated variables into a more concise set of uncorrelated variables, termed principal components (PCs). Each PC represents a linear combination of the original variables. The first principal component captures the maximum possible variance, thus explaining as much variation in the data as possible. The second PC accounts for the second-largest proportion of variance unexplained by the first principal component, and this pattern continues for all subsequent PCs. These PCs, being linear combinations, aim to capture as much of the remaining variation as possible that was not covered by the preceding components.

In our context, each observation, represented by HBV sequences, is portrayed as a linear combination of the first few PCs. A schematic representation of PCA applied to the motif representation data is illustrated in **Fig. 1**. This analytical approach facilitates the reduction of dimensionality, offering a clearer understanding of the underlying patterns in the motif representation data.

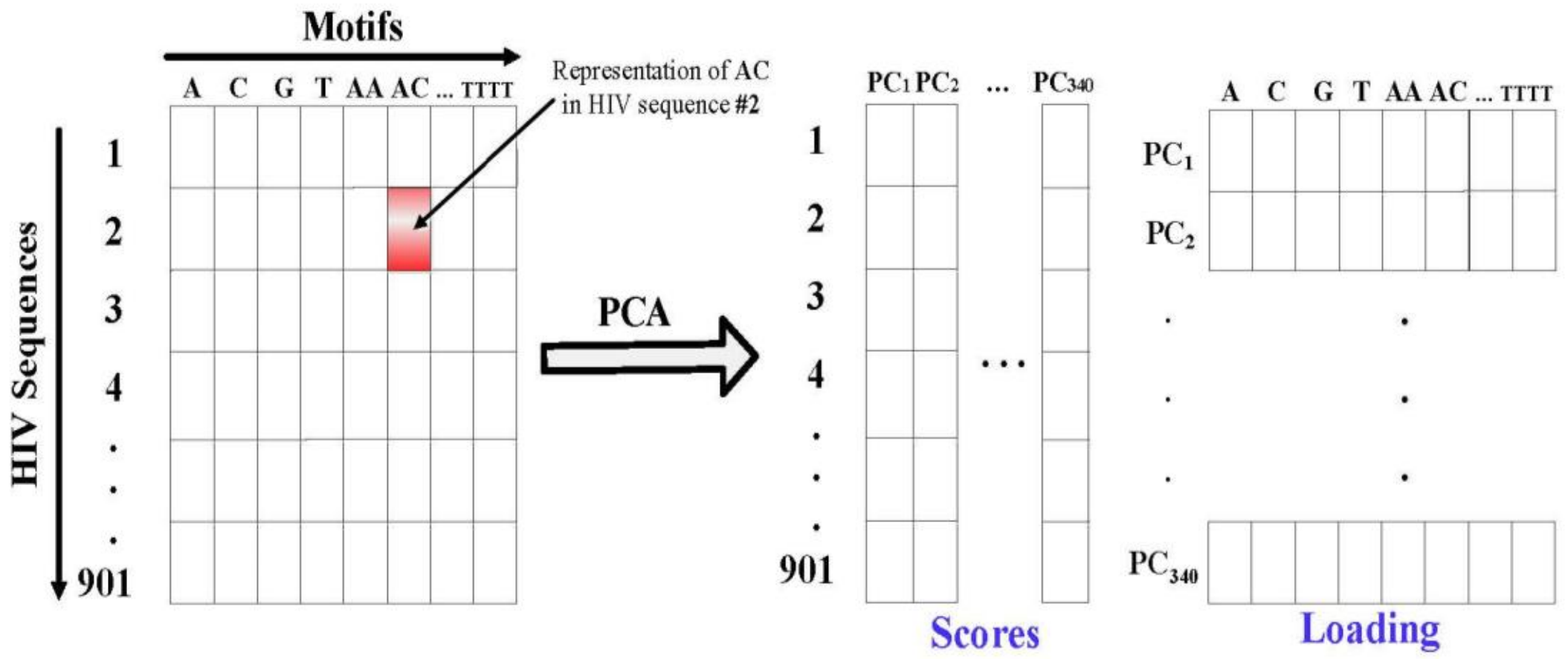

**Figure 1.** A schematic of principal component analysis applied to the motif representation data of HBV sequences. X-axis is 340 k-mers motifs, and Y-axis is 901 HBV sequences.

The PCs to build PCA are calculated as follows. Let's suppose that we have N different observations of a given set of random variables X={$X_1$, $X_2$ … $X_P$} organized in an N×P matrix. Then each PC is defined as a linear regression, predicting $PC_i$ from $X_1$,…, $X_P$:

$$PC_1 = e_{11}X_1 + e_{12}X_2 + \ldots e_{1p}X_p \quad \text{Eq.7}$$

$$PC_2 = e_{21}X1 + e_{22}X_2 + \ldots e_{2p}X_p \quad \text{Eq.8}$$

$$PC_p = e_{p1}X_1 + e_{p2}X_2 + \ldots e_{pp}X_p \quad \text{Eq.9}$$

The coefficients '*e*' can be calculated from eigenvalues and eigenvectors of the covariance matrix of the original dataset [18]. The matrix that contains the $e_{ij}$ coefficients is called the loading matrix, which determines the share of each variable in each PC. Another important matrix is the score matrix '*si*', which allows each observation to be expressed as a linear combination of PCs. For example, the $i^{th}$ observation, $O_i$ can be expressed as:

$$O_i = si_1 \times PC_1 + si_2 \times PC_2 + \ldots + si_p \times PC_p \quad \text{Eq.10}$$

MATLAB has a built-in function for performing PCA, which is in the following format:

[coeff, score, latent, tsquared, explained, mu] = PCA (O); where the input and output parameters are explained in **Table 2**.

In our PCA analysis, the scores matrix (HBV sequences x principal components) describes the relationship (similarity/dissimilarity) between HBV sequences in terms of latent variables (principal components) that are representative of the original variables (i.e., motifs). Likewise, the loadings matrix (principal components x motifs) includes information about the similarity/dissimilarity between motifs in terms of latent variables (principal components) that are representative of the original objects (i.e., HBV sequences).

**Table 2.** Input and output parameters of PCA analysis in Matlab 2017b.

| | Parameter | Description |
|---|---|---|
| 1 | O | The dataset (a matrix of N*P) |
| 2 | coeff | Loading matrix (P*P) |
| 3 | score | Score matrix (N*P) |
| 4 | latent | Eigenvalues of the covariance matrix of X (1*P). |
| 5 | tsquared | Sum of squares of the standardized scores for each observation (1*N) |
| 6 | explained | Percentage of the total variance explained by each principal component (1*P) |
| 7 | mu | Estimated means of the variables in X |

To investigate the grouping of HBV sequences, columns of the scores matrix were plotted against one another. Similarly, by plotting different rows of the loadings matrix against one another, the groupings of motifs were investigated. The data were auto-scaled as a pre-processing step. This was done by subtracting the data in each column from the average of the column of the matrix (i.e., each motif). After this centralization step, we normalized the data by dividing the data in each column by its corresponding standard deviation.

## 3. Results and discussions

To investigate the presence of principal components containing information related to the motif preference of hypermutated sequences, we examined all principal components up to the 20th component.

### Verification of D-Ratio Approach through Identification of Different HIV-1 Subtypes

To assess the effectiveness of the D-Ratio approach, we initially applied the method to HIV-1 sequences obtained from the Los Alamos National Laboratory (LANL) database. The goal was to identify HIV-1 subtypes and discern the motif preferences of APOBEC3 proteins within HIV-1 genomes. The observed subtype differences among the HIV-1 sequences presented a significant source of variation in the motif representation data.

In **Fig. 2a**, the principal component analysis of PC1 vs. PC2 is depicted, illustrating the score plot. This analysis confirmed the capability of the D-ratio method, revealing four distinct clusters of HIV-1 sequences, each corresponding to one of the HIV-1 subtypes. Notably, a striking feature was observed in the cluster scores containing subtype B sequences, indicating a difference from the other clusters (subtypes A1, C, and recombinant 01_AE). Specifically, PC1 exhibited a positive score for all subtype B sequences but negative scores for the other subtypes. Furthermore, PC2 provided additional information about subtypes C, A1, and recombinant 01_AE, segregating clusters into positive and negative score groups.

Previous research has indicated that genomic differences between HIV-1 subtypes are not confined to specific motifs or groups of motifs [3, 4]. Instead, these differences manifest across entire genomes of diverse subtypes. Moreover, as depicted in the loading plot of PC1 vs. PC2 in **Fig. 2b**, our analysis did not reveal distinct clusters of motifs between different HIV-1 subtypes. This observation suggests that the differences among HIV-1 subtypes do not stem from disparate groups of motifs for each subtype.

This verification underscores the robustness of the D-Ratio approach in elucidating not only the subtypes but also the motif preferences within the HIV-1 genomes. The absence of discernible motif clusters between subtypes further supports the notion that genomic differences extend beyond specific motifs, emphasizing the utility of a comprehensive approach to capture the broader genomic distinctions among HIV-1 subtypes.

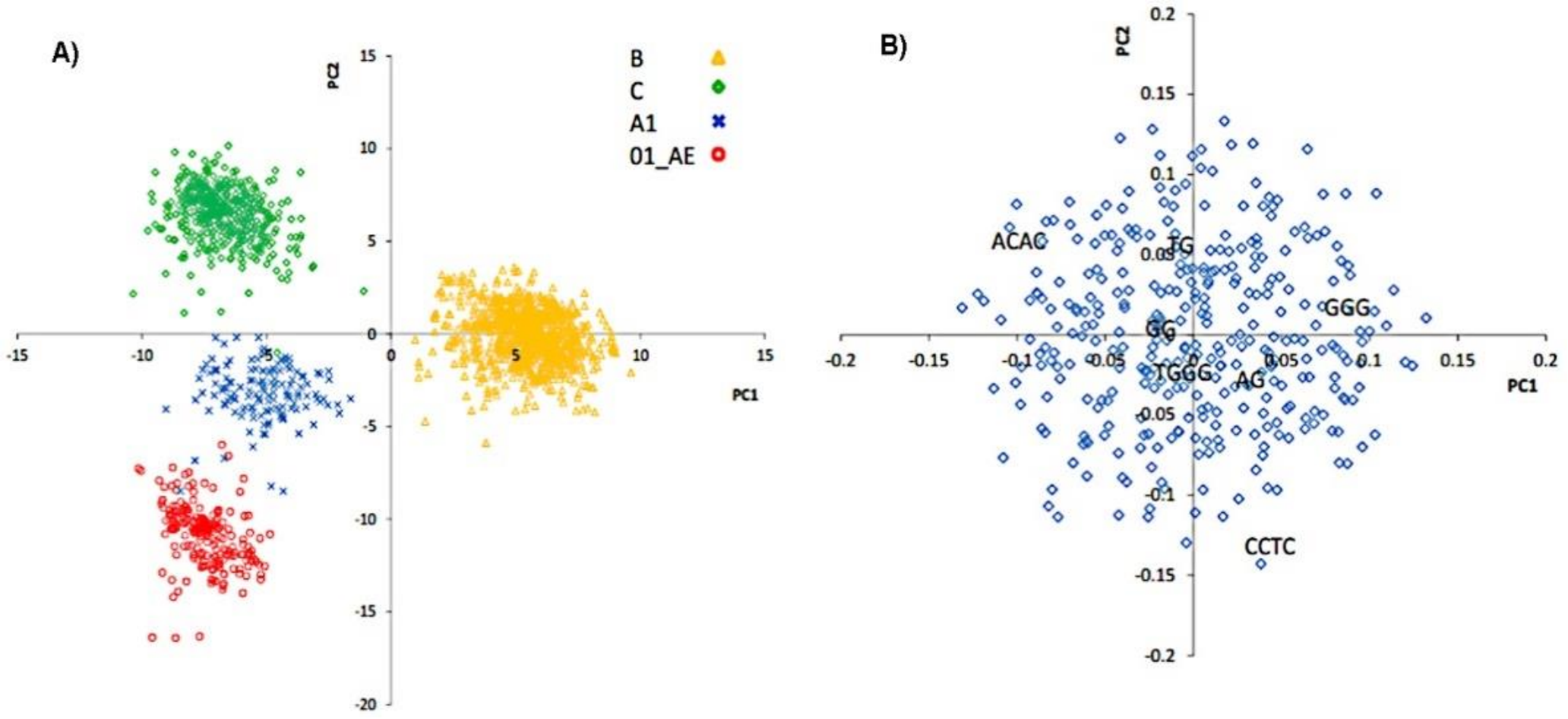

**Figure 2.** Principal component analysis of the motif representation data of HIV-1 sequences: **a)** Scores plot (PC1 *vs.* PC2) of the motif representation data of HIV-1 sequences from subtypes/recombinant B, C, A1, and 01_AE. Each point is an HIV-1 full genome sequence. **b)** Loadings plot (PC1 *vs.* PC2) of the motif representation data of HIV-1 sequences from subtypes/recombinant B, C, A1, and 01_AE. Each point is a k-mers motif. The figure was adopted from the previous work published in PLoS One [22].

**Identification of hypermutation by APOBEC3G in HIV-1 genomes**

While PC1 and PC2 effectively discriminated among HIV-1 subtypes, subsequent principal components (PCs) yielded insightful information regarding the motif preference of APOBEC3G. In **Fig. 3a**, the score plot of PC3 vs. PC4 for HIV-1 sequences pertaining to subtypes A1, C, B, and 01_AE is presented. The plot reveals a primary cluster of sequences at the center and several outlier sequences marked with "H," indicating sequences hypermutated by APOBEC3G (as corroborated by the Los Alamos National Laboratory HIV database). Additionally, two other outlier sequences are marked as hypermutated sequences by APOBEC3F in the LANL HIV database.

**Fig. 3b** illustrates the loading plot of PC3 vs. PC4, mirroring the structure of the score plot. The loading plot exhibits a central cluster of motifs and several outlier motifs arranged in two inverse orientations. The identification of hypermutations in the score plots allows these outlier motifs to be leveraged for discerning the motif preference of APOBEC3G. A well-established fact is that APOBEC3G induces mutations in G within 2-mers GG, which are not flanked by a 3′ C [3, 4]. Notably, target motifs GG, GGG, TGG, and TGGG, favored by APOBEC3G, emerge prominently at the top right-hand side of the loading plot. In the opposite orientation, product motifs AG, AGG, TAG, and TAGG are evident in the bottom left-hand side of the loading plot. Additionally, motifs GC, TGC, and TGGC are observed within the product motifs, representing disfavored target motifs. Further discussion on why these motifs are among the product motifs will be expounded upon in the subsequent discussion section.

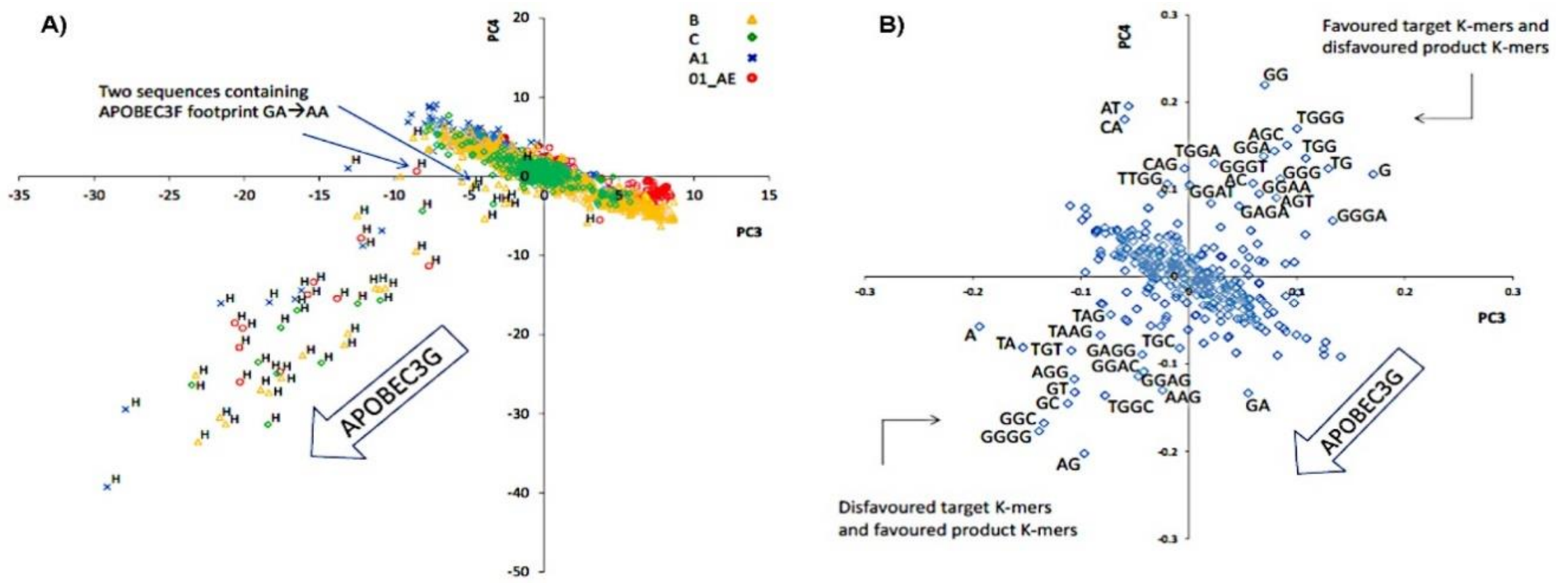

**Figure 3.** Principal component analysis of the motif representation data of HIV-1 sequences. **a)** Scores plot (PC3 *vs.* PC4) of the motif representation data of HIV-1 sequences from subtypes/recombinant B, C, A1, and 01_AE. Each point is an HIV-1 full genome. HIV-1 from subtypes B, C, A1, and the recombinant 01_AE are shown by orange, green, blue, and red, respectively and hypermutated sequences recognized by the Los Alamos HIV Sequence Database are indicated by ''H''. **b)** Loading plot (PC3 *vs.* PC4) of the motif representation data of HIV-1 sequences from subtypes/recombinant B, C, A1 and 01_AE. Each point is a k-mers motif. The motifs related to hypermutation by APOBEC3G appear as outliers. The figure has been adapted from our previous work published in PLoS One [22].

Utilizing established HIV-1 hypermutated sequences, we demonstrate the efficacy of the D-Ratio approach in identifying APOBEC3 protein footprints without requiring alignment. In the subsequent sections, we aim to employ the D-Ratio approach to explore potential motif preferences of APOBEC3 proteins within the HBV genome.

#### Comparison of Normal HBV Sequences with Hypermutated HBV Sequences

To assess the impact of motifs in representing our data, we initially applied Principal Component Analysis to all HBV sequences collectively, seeking to discern differences between normal and hypermutated HBV sequences. Our findings indicate that subtype differences among the HBV sequences constitute a significant source of variation in the motif representation data.

As depicted in Fig. 4a, the principal component analysis of PC1 vs. PC2 confirms the effectiveness of our method, revealing two primary clusters of HBV sequences, encompassing normal and likely hypermutated HBV sequences. Conversely, **Fig. 4b** illustrates the scores of PC1 vs. PC2 for the first study involving 3000 HBV sequences, where each cluster corresponds to one of the HBV subtypes.

To investigate the motif preferences of APOBEC3G and APOBEC3nonG, we subsequently apply PCA specifically to 901 likely HBV hypermutated sequences. This focused analysis aims to uncover potential patterns in motif representation related to hypermutation by these APOBEC3 enzymes.

Using the known HIV-1 hypermutated sequences, we show that the D-Ratio approach can identify APOBEC3 protein footprint without needing any alignment. In the next sections, we want to use D-Ratio approach to investigate if APOBEC3 proteins have any preference motif in the HBV genome or not.

*Comparison of normal HBV sequences with hypermutated HBV sequences*

To investigate the impact of motifs to represent our data, we first apply the PCA to all HBV sequences together to find the difference between normal and hypermutated HBV sequences. Our results show that the subtype differences among the HBV sequences result in a remarkable source of variation in the motif representation data.

As is shown in **Fig. 4a** the principal component analysis of PC1 *vs.* PC2 confirmed the ability of our method and showed two main clusters of HBV sequences, including HBV normal sequences and HBV likely hypermutated sequences. On the other hand, **Fig. 4b** shows the scores of PC1 vs. PC2 for only the first study (3000 HBV sequences) where each cluster includes one of the HBV subtypes.

To investigate the motif preferences of APOBEC3G and APOBEC3nonG, we apply PCA on 901 likely HBV hypermutated sequences.

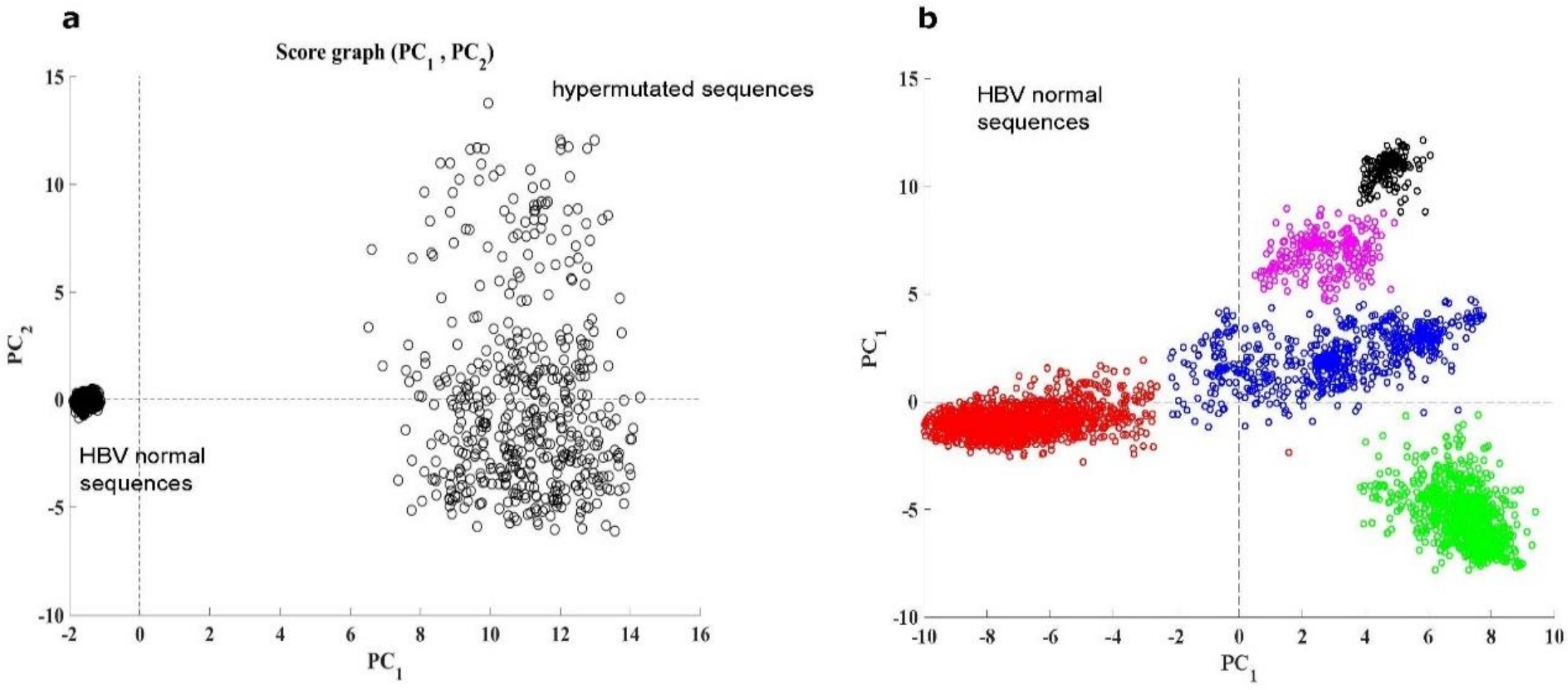

**Figure 4.** Principal component analysis of the motif representation data of HBV sequences: **a)** Scores plot (PC1 *vs.* PC2) of the motif representation data of 3000 HBV normal sequences from NCBI and 901 HBV hypermutated sequences from Vartanian *et al*. [23]. Each point is a HBV sequence**. b)** Scores plot (PC1 *vs.* PC2) of the motif representation data of only 3000 HBV normal sequences.

**Identification of Hypermutation by APOBEC3 Enzymes**

In order to discern the motif preferences of APOBEC3G and APOBEC3nonG, Principal Component Analysis was applied to a dataset consisting of 901 hypermutated HBV sequences. **Fig. 5a** illustrates the score plot of PC1 vs. PC2 for these 901 HBV sequences, where each data point represents a distinct HBV sequence. **Fig. 5b** provides the corresponding loading plot of PC1 vs. PC2. The plot indicates a primary cluster of sequences at the center, alongside several outlier sequences that appear to be potential hypermutated sequences. A detailed manual investigation of these outliers confirmed the presence of numerous mutations. A zoomed-in view of the loading plot related to PC1 vs. PC2 is presented in **Fig. 5c**.

Leveraging these outlier hypermutated sequences allows for an exploration of the motif preference of APOBEC3G or APOBEC3nonG in HBV hypermutated genomes. Previous studies have reported that APOBEC3G predominantly induces mutations in G within 2-mers GG, not flanked by a 3′ C. Target motifs such as GG, GGG, TGG, and TGGG are particularly favored by APOBEC3G [1, 4]. Similarly, it has been documented that APOBEC3nonG, primarily APOBEC3F, induces mutations in G within 2-mers GA not flanked by a 3′ C, with preferred target motifs being GA, TGA, and TGGA [1, 4].

In **Fig. 5b**, the loading plot reveals that the target motif TGGG is prominent at the bottom right-hand side, while product motifs TGAG emerge on the top left-hand side. Additionally, motifs of TGGC are observed within the product motifs, representing disfavored target motifs. Notably, the 4-mer TGGC, although not classified as a product motif, appears as an overrepresented motif due to hypermutation, as the frequencies of GGC do not change. This observation highlights the nuanced impact of hypermutation on motif representation within the HBV genome.

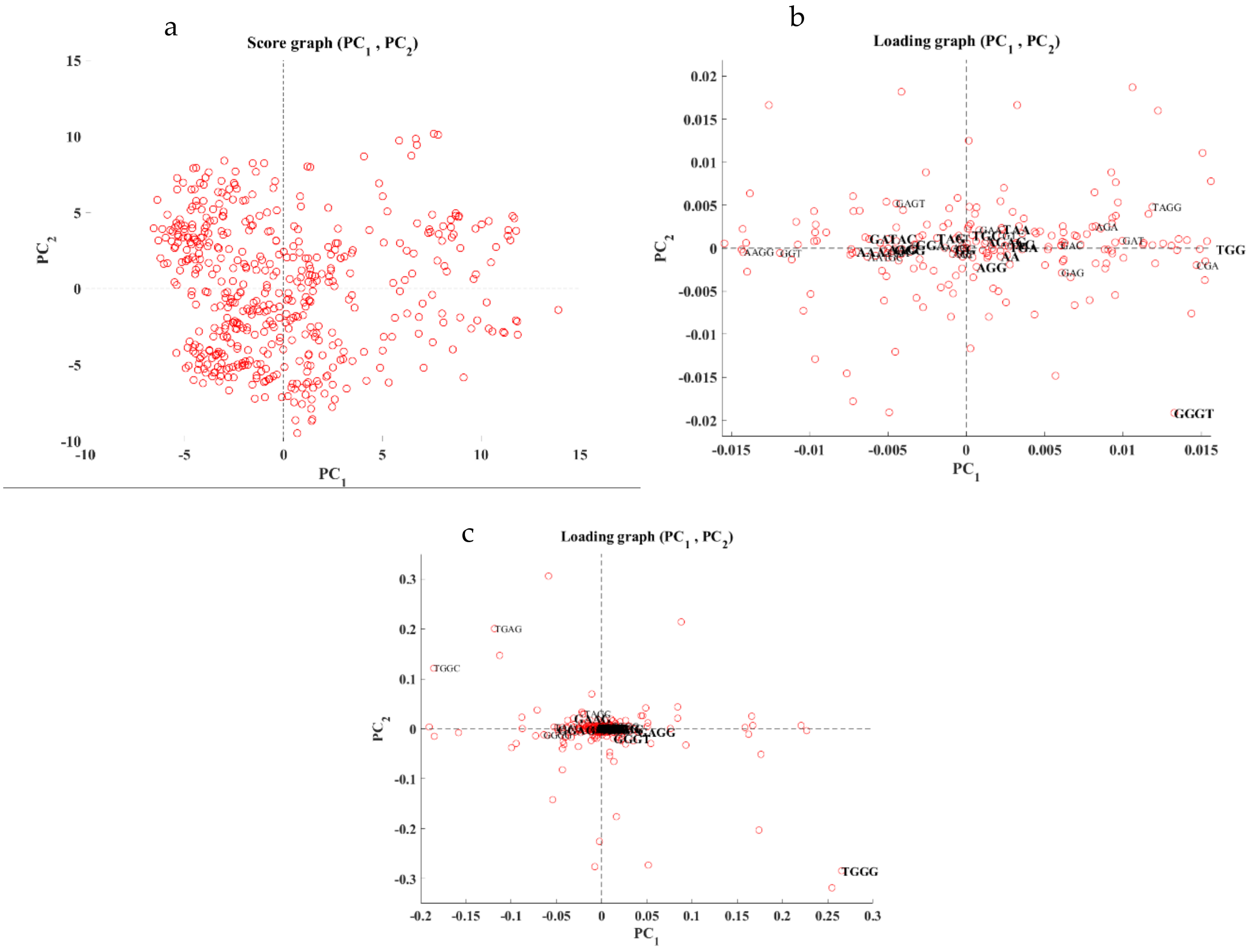

**Fig.5.** Principal component analysis of the motif representation of 901 hypermutated HBV sequences. **a)** Scores plot (PC1 *vs.* PC2) of the motif representation data of 901 hypermutated HBV sequences. Each point is a HBV sequence. **b)** Loading plot PC1 *vs.* PC2 in the 901 hypermutated HBV sequences. Each point is a motif. **c)** A zoom view of loading plot PC1 *vs.* PC2 in the 901 hypermutated HBV sequences.

In **Fig. 5c**, a detailed examination of the zoomed-in view of the loading plot from **Fig. 5b** reveals distinct orientations for the target motif GGGT and the product motif GAGT. Notably, motifs TAGC and GAAG are identified in the product section of the plot, while no corresponding target motifs (likely TGGC and GGGG, respectively) are observed in the target section of the plot.

Expanding our investigation to encompass all principal components up to PC19 vs. PC20, with a focus on GA and AA 2-mer motifs as outlier motifs in two inverse orientations, yielded **Fig. 6.** However, examination of the loading plots in **Fig. 6** did not reveal sensible target and product motifs in the two opposite orientations. Despite the expectation that motifs GA and AA, as well as GG and AG, should serve as target and product motifs for APOBEC3nonG and APOBEC3G, respectively, these motifs did not manifest in the loading plots across different PCs in **Fig. 6.** Furthermore, no other outlier motifs emerged that could be identified as favored target and product motifs of APOBEC3G and APOBEC3nonG. In the loading plot PC7 vs. PC8, 2-mer motifs GA and AA were detected but in the same orientations.

Contrary to expectations, well-established target motifs for APOBEC3G (GG, GGG, GGGG, TGG, and TGGG) and APOBEC3nonG (GA, TGA, and TGGA) in HIV sequences [1, 4, 22] did not appear as outliers in the loading plots. This raises questions about the transferability of these motifs to the context of HBV hypermutated sequences. Further elucidation of the motif preferences of APOBEC3G and APOBEC3nonG in the specific context of HBV genomes requires additional investigation and analysis.

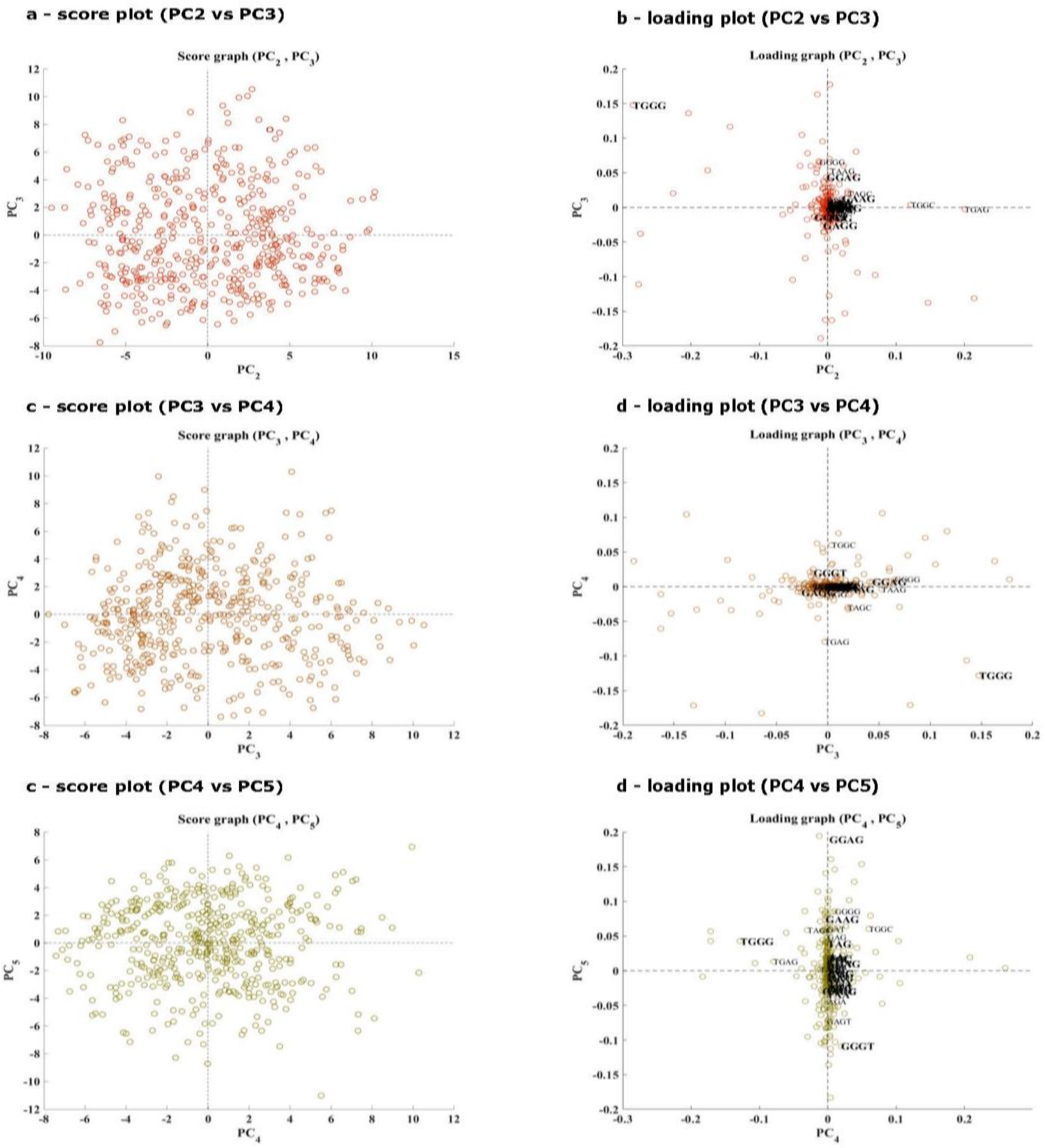

**Figure 6.** Principal component analysis of the motif representation data of 901 HBV sequences. **a)** Scores plot (PC2 *vs.* PC3) of the motif representation data of 901 hypermutated HBV sequences. Each point is a HBV sequence. **b)** Loading plot (PC2 *vs.* PC3) in the 901 hypermutated HBV sequences. Each point is a motif. **c)** Scores plot (PC3 *vs.* PC4) of the motif representation data of 901 hypermutated HBV sequences. **d)** Loading plot (PC3 *vs.* PC4) in the 901 hypermutated HBV sequences. **e)** Scores plot (PC4 *vs.* PC5) of the motif representation data of 901 hypermutated HBV sequences. **f)** Loading plot (PC4 *vs.* PC5) in the 901 hypermutated HBV sequences.

In the multivariate exploratory analysis of motif representation data, a notable revelation emerged—contrary to prior reports, the recognized motif preferences of APOBEC3G and APOBEC3nonG in the HBV genome did not manifest as favored target sites for these enzymes. This finding challenges the conventional understanding that APOBEC3G induces G-to-A mutations within GG motifs that lack flanking Cs at both the first and second 3′ positions (+1 and +2). Surprisingly, our analysis demonstrated that no GG motif emerged as an outlier in the HBV hypermutated sequences.

As discussed in the introduction, numerous studies have previously reported the presence of APOBEC3 enzyme footprints on the HBV genome. However, our meticulous investigation into the footprints of APOBEC3G and APOBEC3nonG fails to corroborate these earlier results. Contrary to existing literature, our findings do not lend support to the existence of any discernible motif preference for APOBEC3 enzymes within the HBV genome.

This unexpected outcome underscores the complexity and context-specific nature of APOBEC3 interactions with viral genomes, emphasizing the need for nuanced investigations tailored to specific viral contexts. Further research is imperative to elucidate the distinctive mechanisms governing APOBEC3 activity within the HBV genome and to refine our understanding of the interplay between APOBEC3 enzymes and different viral substrates.

## 4. Conclusions

The application of machine learning methods in medical data analytics enables the extraction of valuable insights, prediction of patient outcomes [24, 25], and personalized healthcare interventions, revolutionizing the field by harnessing the power of data to improve diagnostics and treatment strategies [26-27]. In our extensive motif analysis aimed at unraveling the APOBEC3 enzyme footprints on HBV sequences, intriguing findings have emerged. Specifically, our investigation identified only two target motifs, TGGG and GGGT, as apparent preferences for APOBEC3G within the HBV genome. However, the anticipated motifs GG, TGG, GGG, GGGG, and GA, TGA, and GAA—previously established as favored motifs for APOBEC3G and APOBEC3nonG in other contexts—failed to manifest as preference motifs in our analyses.

These outcomes raise several compelling questions regarding the interaction of APOBEC3 enzymes with the HBV genome. One interpretation is that APOBEC3G and APOBEC3nonG may not possess distinct motif preferences for inducing G-to-A mutations in the HBV viral sequences. Alternatively, these APOBEC3 enzymes might not be actively involved in the mutational processes within the HBV genome. Nevertheless, the observed patterns could also be attributed to the constraints imposed by the limited available data.

Delving into the distinct mechanisms governing the activities of APOBEC3G and APOBEC3nonG provides further insight into these unexpected findings. Previous studies have highlighted the unique preferences of these enzymes in inducing mutations in viral genomes.

APOBEC3G, for instance, has been widely recognized for its ability to induce G-to-A mutations within GG motifs, specifically when not flanked by a 3′ C at both the +1 and +2 positions. This characteristic has been attributed to the enzyme's deaminase activity, leading to hypermutation in target motifs [1, 4, 22]. However, our study challenges this paradigm by revealing the absence of GG motifs as preference motifs for APOBEC3G in the context of HBV hypermutated sequences.

On the other hand, APOBEC3nonG, primarily APOBEC3F, has been reported to favor G-to-A mutations within 2-mers GA that lack a 3′ C. Target motifs GA, TGA, and TGGA have been identified as preferred motifs for APOBEC3nonG in previous investigations [1, 11, 20]. Strikingly, our findings did not confirm the presence of these motifs as preference motifs for APOBEC3nonG in the context of HBV.

These disparities may indicate unique regulatory mechanisms or contextual variations influencing the activities of APOBEC3G and APOBEC3nonG in the HBV genome. For instance, it is plausible that the distinct genomic architecture of HBV influences the binding and targeting preferences of these enzymes. Moreover, potential interactions with host factors or viral proteins specific to the HBV context could contribute to the observed differences.

In conclusion, our study underscores the complexity of APOBEC3 interactions with viral genomes and emphasizes the need for nuanced investigations tailored to specific viral contexts. The unexpected findings open avenues for further research, necessitating larger datasets, diverse viral contexts, and experimental validations to unravel the intricate dynamics between host APOBEC3 enzymes and the HBV genome. Such insights hold promising implications for antiviral strategies and therapeutic interventions, warranting continued exploration into the intricacies of host-virus interactions.

**Author Contributions:** KJ developed the concept. KJ and ES developed the method. KJ, ES, and ID wrote the manuscript. KJ, MS, ID, RA revised the manuscript. KJ, ES, and MS analyzed the data. All authors read and approved the final manuscript.

**Funding:** NA.

**Acknowledgments:** We thank Professor Jean-Pierre Vartanian for providing us with HBV viral sequences. The source code of the proposed method and NCBI sequences are available upon request.

**Conflicts of Interest:** The authors declare no conflict of interest.